\documentclass[prd,twocolumn,superscriptaddress,nofootinbib,nobalancelastpage
  ,noshowpacs,preprintnumbers,longbibliography
  ]{revtex4-1}
\usepackage{amsmath,amssymb,mhchem}
\usepackage{graphicx}
\usepackage{url}
\usepackage{color}
\usepackage[dvipsnames]{xcolor}
\usepackage[colorlinks=true,breaklinks=true]{hyperref}
\hypersetup{allcolors=[rgb]{0.0 0.0 0.6},linkcolor=[rgb]{0.75 0.05 0.05}}
\usepackage{float}

\newcommand{\exclude}[1]{}

\begin{document}

\preprint{MPP-2018-213}

\title{Cosmic infrared background excess from axionlike particles and implications for multimessenger observations of blazars
}

\author{Oleg E. Kalashev}
\email[]{kalashev@inr.ac.ru}

\affiliation{Institute for Nuclear Research, 60th October Anniversary Prospect 7a, Moscow 117312 Russia}
\affiliation{Moscow Institute for Physics and Technology, 9 Institutskiy per., Dolgoprudny, Moscow Region, 141701 Russia}

\author{Alexander Kusenko}
\email[]{kusenko@ucla.edu}

\affiliation{Department of Physics and Astronomy, University of California, Los Angeles, CA 90095-1547, USA}
\affiliation{Kavli IPMU (WPI), University of Tokyo, Kashiwa, Chiba 277-8
568, Japan}

\author{Edoardo Vitagliano}
\email[]{edovita@mpp.mpg.de}

\affiliation{Max-Planck-Institut f\"ur Physik (Werner-Heisenberg-Institut), F\"ohringer Ring 6, 80805 M\"unchen, Germany}



\date{\today}

\begin{abstract}
The first measurement of the diffuse background spectrum at 0.8-1.7 $\mu \rm{m}$ from the CIBER experiment has revealed a significant excess of the cosmic infrared background (CIB) radiation compared to the theoretically expected spectrum.  We revisit  the hypothesis that decays of axionlike particle (ALP) can explain this excess, extending previous analyses to the case of a warm relic population.  We show that such a scenario is not excluded by anisotropy measurements nor by stellar cooling arguments. Moreover, we find that the increased extragalactic background light (EBL) does not contradict  observations of blazar spectra.
Furthermore, the increased EBL attenuates the diffuse TeV gamma-ray flux and  alleviates the tension  between the detected neutrino and gamma ray fluxes.
\end{abstract}


\maketitle

\section{Introduction}
 
 Recently, the Cosmic Infrared Background Experiment (CIBER) collaboration has claimed the detection of an unexpectedly high flux compared to theoretical expectations in the 0.8-1.7 $\mu \rm{m}$ range of wavelengths \cite{Matsuura:2017lub}. This measurement is complementary to other observations in the infrared band like the ones carried by AKARI \cite{Seo:2015fga} and IRTS \cite{Matsumoto:2015fma}. Even if an astrophysical explanation of the detected excess or systematic errors are not ruled out, it is worthwhile to speculate about a possible  flux due to big bang relics, such as an axionlike particles (ALP) with mass around 1 eV.  ALPs generalize the concept of the axion, introduced to solve the so-called Strong CP problem, which has multifaceted phenomenology \cite{Marsh:2015xka}. However, ALPs could have coupling to particles besides the one to the photon, e.g. involving a hidden photon. The contribution of such ALP decays to the cosmic infrared background (CIB) was examined in Ref.~\citep{Kohri:2017oqn}.  Here we will revisit the hypothesis, taking into account the detector energy resolution,  the possibility of warm dark matter, and the implications of increased EBL for blazar multimessenger observations.

While a solid lower bound to the CIB radiation can be obtained through deep sky galaxy counts \cite{Keenan:2010na}, the precise shape and intensity of the diffuse, unresolved spectrum in the near-infrared wavelength range is still unknown. 
Direct measurements~\cite{Bock:2005sy,Tsumura:2013iza,Seo:2015fga} are difficult because of the large uncertainties caused by zodiacal light.  Theoretical models are also subject to uncertainties, which result in different predictions~\cite{Gilmore:2011ks,Inoue:2012bk,Scully:2014wpa}. The uncertainties make it difficult to identify any additional contribution to the extragalactic background light (EBL) besides the standard flux due to galaxy emission.  Possible enhancements could come from ultraviolet redshifted photons produced by bottom-up astrophysical accelerators, ranging from high redshift galaxies \cite{Yue:2012dd} to black holes \cite{Yue:2013hya}. 

The EBL can be also measured indirectly. Very high energy gamma rays from blazars have been used to set an upper limits on infrared background radiation~\cite{Costamante:2013sva}. An indirect measurement has been recently carried out using 739 active galaxies and one gamma-ray burst~\cite{Ajello:2018sxm}. However, while such kind of measurements could in principle strongly constrain substantial contributions not resolved by deep galaxy surveys, the possibility of secondary gamma rays produced by cosmic rays along the line of sight~\cite{Essey:2009ju,Essey:2009zg,Essey:2010er,Murase:2011cy,Aharonian:2012fu} undermine these upper bounds. 

In the last few years, searches for indirect probes of portals connecting the standard model of particle physics with the dark matter sector have been pursued (see, e.g., \cite{Gaskins:2016cha}). ALPs as a dark matter candidate have recently received great attention due to the non detection of weakly interactive massive particles \cite{Irastorza:2018dyq}.  It is, therefore, important to examine the CIB data in light of the ALP hypothesis. 

Apart from the increase of the EBL, this hypothesis has an observable impact on the propagation of TeV photons because it implies an enhanced opaqueness through $\gamma \gamma \rightarrow e^+ e^-$ processes. A higher level of EBL would help alleviate the tension between the observed neutrino spectrum and the gamma-ray spectrum of blazars, as discussed below.  We present a case study in which multimessenger, multiwavelength observations can be exploited to obtain new tools to indirectly probe fundamental physics beyond the standard model, making use of data from neutrino telescopes (IceCube), gamma-ray satellites (Fermi-LAT) and sounding rockets equipped with infrared cameras (CIBER), extending the already flourishing multimessenger astronomy tools \cite{IceCube:2018dnn}.

The paper is organized as follows. We will review the CIBER data and the particle physics content of an ALP model; we will then tackle the bounds coming from anisotropy observations by the Hubble Space Telescope and CIBER itself. Later we show how the increase of the CIB affects the propagation of ultra-TeV gamma-rays. This brings us to a final discussion and our conclusions.\label{intro}

\label{fluxsec}
\section{Flux from axionlike particle decay}

\begin{figure*}
\centering
        \includegraphics[width=0.35\textwidth]{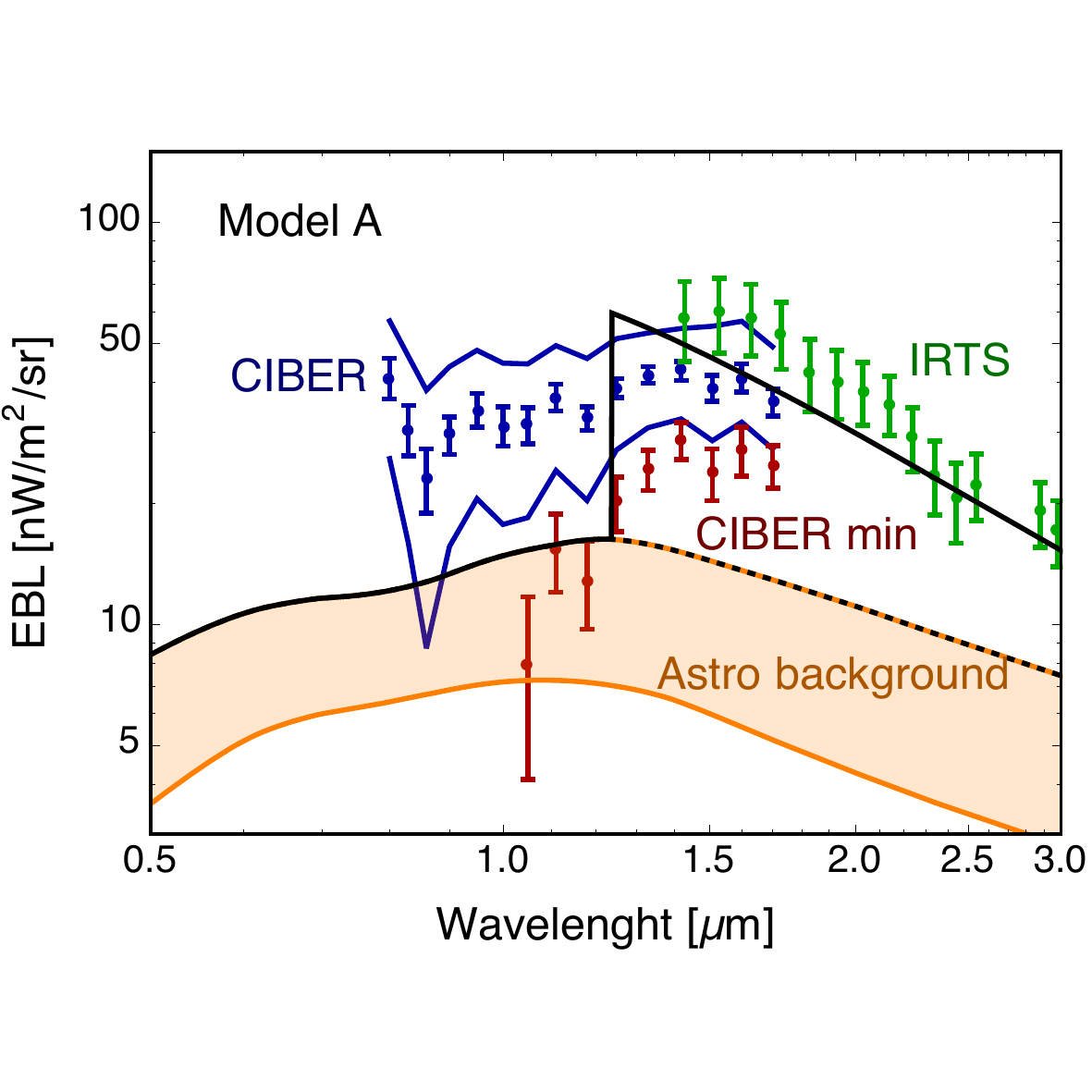}
        \includegraphics[width=0.35\textwidth]{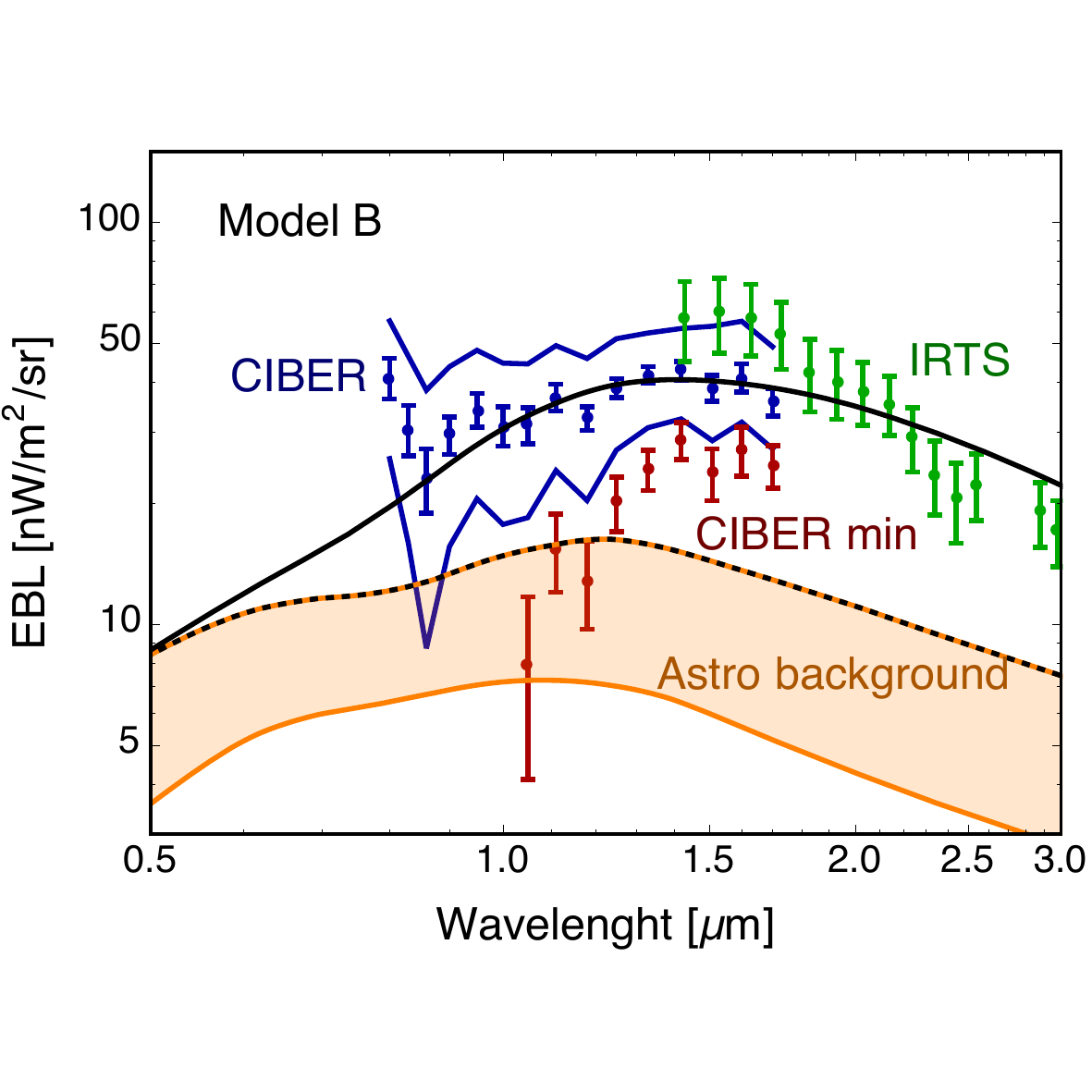}
           \vskip -2em
  \caption{Photon intensity flux from a decaying cold dark matter ALP. Experimental data include CIBER data with Kelsall ZL model (blue, continuous lines are systematic error), CIBER with minimum EBL model (red), IRTS (green) \cite{Matsuura:2017lub,Matsumoto:2015fma}. The total flux (solid black) include the flux from ALP decay and the astrophysical diffuse (dotted black), which we assume to be the upper bound of the band reported in \cite{Stecker:2016fsg}, shown in orange. \textit{Left, model A}: $\omega_{\rm max}=1 \rm \, eV$, $\tau=2\times 10^{22} \rm \, s$, $R=2/3$.  \textit{Right, model B}:  $\omega_{\rm max}=8 \rm \, eV$, $\tau=1\times 10^{16} \rm \, s$, $R=2\times10^{-4}$.
  }
  \label{fig:alpdecay}
\end{figure*}

\begin{figure*}
\centering
            \includegraphics[width=0.35\textwidth]{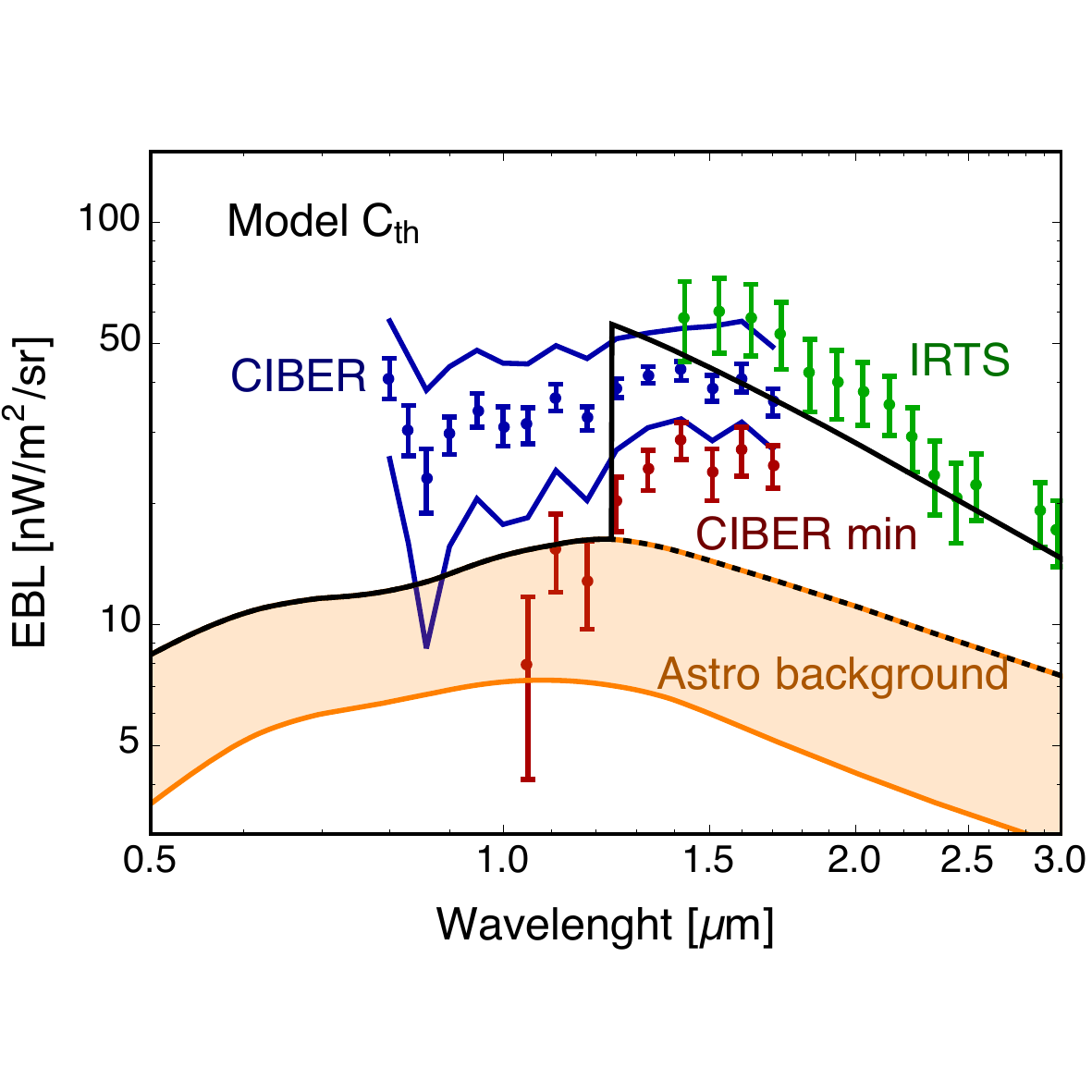}
        \includegraphics[width=0.35\textwidth]{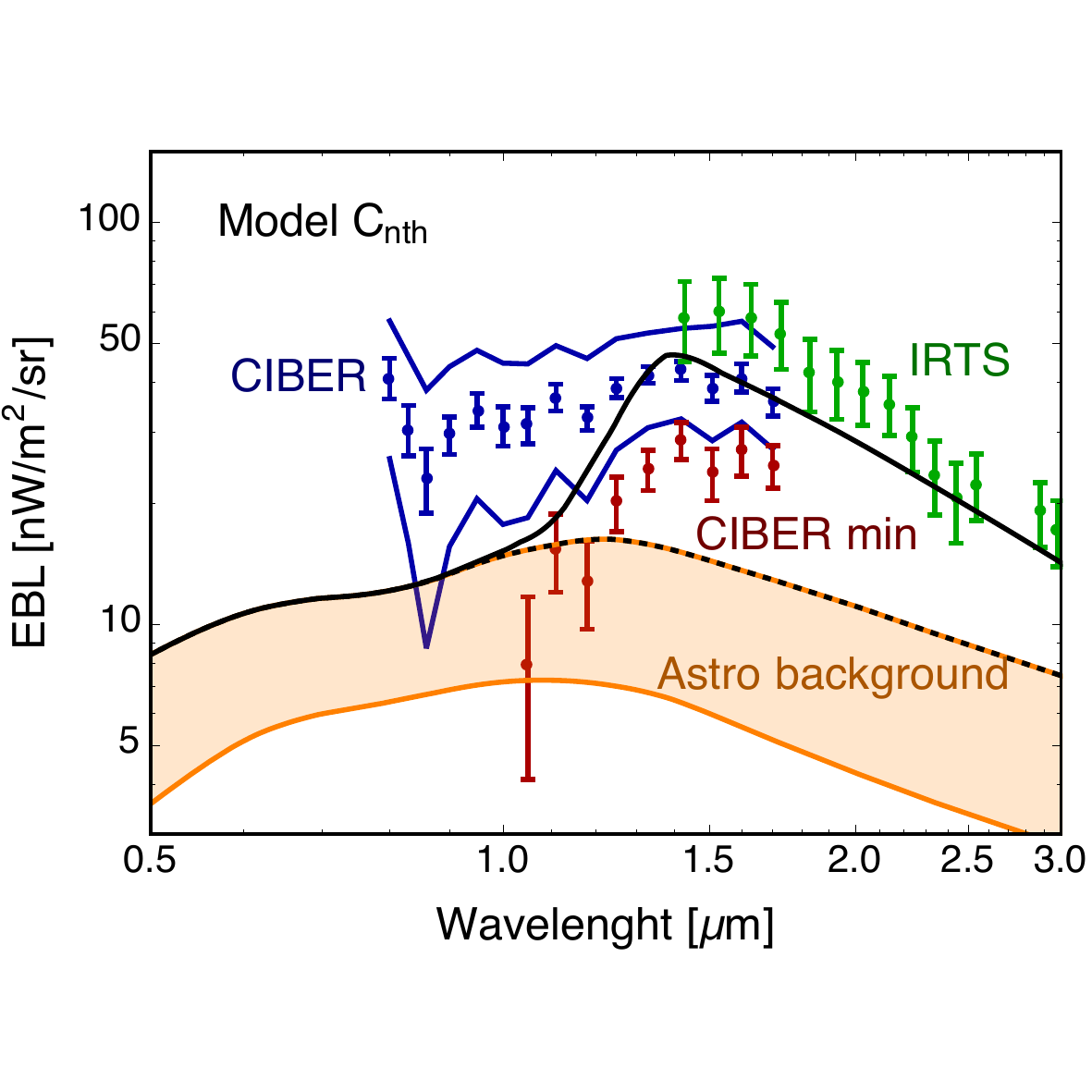}
           \vskip -2em
  \caption{Same as Fig.~\ref{fig:alpdecay}, but assuming a decaying warm dark matter ALP. \textit{Left, model $\rm C_{th}$:} $T_{\rm th}=0.37 \, T_\gamma^{(0)}=0.086 \rm{\, meV}$, $\omega_{\rm max}=1 \rm \, eV$, $\tau=9\times 10^{21} \rm \, s$, $R_{\rm th}=7\times 10^{-3}$. \textit{Right, model $\rm C_{nth}$}:  $T_{\rm nth}=16.7 \rm{\, meV}$, $\omega_{\rm max}=1 \rm \, eV$, $\tau=3\times 10^{21} \rm \, s$, $R_{\rm nth}=2/3\times 10^{-3}$.
  }
  \label{fig:alpdecaywarm}
\end{figure*}

We are interested in the redshift evolution of the diffuse infrared radiation produced by the decay of a relic axionlike particle to a photon and a hidden photon, $a\rightarrow \gamma + \chi$ \cite{Daido:2018dmu, Kohri:2017oqn}. The possibility of having axions with suppressed two-photons coupling has received some attention recently due to the peculiar phenomenology of photophobic axions \cite{Craig:2018kne}. The decay is due to the Chern-Simons~\cite{Chern:1974ft} interaction Lagrangian
\begin{equation}
\mathcal{L}\supset\frac{g_{a\chi\gamma}}{4}a F^{\mu\nu}\tilde{F}^\chi_{\mu\nu}
\end{equation}
where $\tilde{F}^{\mu\nu}=\epsilon_{\mu\nu\rho\sigma}F^{\mu\nu}/2$. While such a coupling between dark matter and photons is not directly inspired by solutions to other problems (like the QCD axion), experimental signatures would be quite different from the ones of the QCD axion, motivating us to explore this class of parametric models. The nonrelativistic decay rate for the ALP is found to be
\begin{equation}
\Gamma=\frac{1}{16\pi}\overline{|\mathcal{M}|^2}\frac{m_a^2-m_\chi^2}{m_a^3}=\frac{g_{a\chi\gamma}^2}{128\pi}\frac{(m_a^2-m_\chi^2)^3}{m_a^3}
\end{equation}
where the squared amplitude averaged over final polarization states is $\overline{|\mathcal{M}|^2}=g_{a\chi\gamma}^2(m_a^2-m_\chi^2)/8$. This correctly reduces to the usual axion decay rate when $m_\chi=0$ and one includes a factor of 2 due to the final state involving identical photons \cite{Cadamuro:2010cz}. Interestingly, the decay rate depends just on one kinematic quantity in the nonrelativistic approximation, namely, the maximum available energy for the outgoing photon 
\begin{equation}
\omega_{\rm max}=\frac{m_a^2-m_\chi^2}{2m_a} \ .
\end{equation} The degeneracy would be broken if the ALP were non-negligibly relativistic.

The energy intensity (energy flux per unit of energy, time, surface per steradians) is computed from a window function $W(z',\omega')$,
\begin{align}\label{eq:enintensity}
 I(\omega)&= \ \frac{\omega^2}{4\pi}\frac{dN}{dS d\omega dt}=\omega^2 \int_z^{\infty}dz' W(z',\omega')   \nonumber \\  &=  \frac{\omega^2}{4\pi}\int_z^\infty\frac{dz'}{H(z')}\frac{(1+z)^2}{(1+z')^3} e^{-\Gamma t(z')}
\nonumber  \\&\times
\int \frac{d^3 \bold{p}_a'}{(2\pi)^3 2 E_a'}\frac{d^3 \bold{p}_\chi'}{(2\pi)^3 2 E_\chi'}\frac{ \omega'}{4\pi^2}
\nonumber \\& \times(2\pi)^4 \delta^{(4)}(p_\chi'+k'-p_a')\overline{|\mathcal{M}|^2}f_a(\bold{p}_a') \ ,
\end{align}
where $H(z)=H^{(0)}\sqrt{\Omega_\Lambda + 
 \Omega_m(1 + z)^3}$ is the Hubble function, $z$ is the redshift at which the flux is ``observed'', $z'$ is the redshift at which $a$ decays with a squared amplitude $\overline{|\mathcal{M}|^2}$, the momentum at the production point is $\omega'=\omega (1+z')/(1+z)=\omega^{(0)} (1+z')$ (as well as $\bold{p}'=\bold{p} (1+z')/(1+z)$), $f_a(\bold{p}_a)$ is the momentum distribution of the ALPs, so that the number density (when there is no decay) is $n_a=\int d^3 \bold{p}_a/(2\pi)^3 f_a(\bold{p}_a)$. In the following the superscript ${(0)} $ will indicate comoving quantities. We include the reduction in the number density due to decay with rate $\Gamma$ over the cosmic time \begin{equation}
t(z')=\frac{1}{3H^{(0)}\sqrt{\Omega_\Lambda}}\log \frac{ \sqrt{\Omega_\Lambda+\Omega_m(1+z')^3}+\sqrt{\Omega_\Lambda}}{\sqrt{\Omega_\Lambda+\Omega_m(1+z')^3}+\sqrt{\Omega_\Lambda}} \ ,
\end{equation}
whereas we do not need to account for absorption; the latter is negligible in the wavelength range under study. The only relevant process reducing the flux of a single source is due to Thomson scattering \cite{1989ApJ...344..551Z}. However, Thomson scattering preserves the energy of the scattering photon. As such, it is irrelevant in the case of diffuse production with no sensible fluctuations in the electron spacial distribution, which we consider in first approximation to be homogeneous. We will now explore two main scenarios, involving cold dark matter or warm dark matter.

\subsection{ALP cold dark matter}
Equation \eqref{eq:enintensity} correctly reduces to Equation (50) of \cite{Cirelli:2010xx}, when one takes a cold dark matter (CDM) distribution for the ALP population, $f_a(\bold{p}_a)=n_a^{(0)} (2\pi)^3\delta^{(3)}(\bold{p}_a) (\bold{p}_a/\bold{p}_a^{(0)})^3$, and gets rid of Dirac deltas. Integration over $z'$ yields
\begin{align}
I(\omega)
&=\frac{1}{4\pi}\omega^2(1+z)^2 n^{(0)}_a \Gamma\int_z^{\infty}dz'  e^{-\Gamma t(z')}\frac{\delta[\omega'-\omega_{\rm max}]}{H(z')} \nonumber \\
&=\frac{1}{4\pi}\omega(1+z)^3 n^{(0)}_a  \Gamma \ e^{-\Gamma t(\tilde{z})}\frac{\theta[\tilde{z}-z]}{H(\tilde{z})}  \ ;
\end{align}
here, $\tilde{z}=(1+z)\omega_{\rm max}/\omega-1$ and $\theta[\tilde{z}-z]$ is the Heaviside function. As expected, the comoving intensity is simply found by multiplying times a $(1+z)^{-4}$ factor (one power coming from $\omega$). For $z=0$ this agrees with Equation (3) of \cite{Kohri:2017oqn}. In the same paper, the (three-fold) parameter space to explain the CIBER excess has been explored. The maximum available energy must be $\omega_{\rm max}\lesssim 10.2 \rm{\, eV}$ to avoid constraints due to reionization and more stringently to the Lyman-alpha forest absorption spectrum. The lifetime $\Gamma^{-1}$ should be roughly of the order of the age of the Universe, and cannot be too small because ALPs can be produced in astrophysical systems, modifying the stellar evolution \cite{Kohri:2017oqn}. 

On the one hand, the ALP decay to a photon plus a hidden photon avoids the direct detection bounds on the coupling $g_{a\chi\gamma}$, which instead constraint the $g_{a\gamma\gamma}$ of standard ALPs (decaying to two photons),\footnote{In principle there should be also the operator $\mathcal{L}\supset\frac{g_{a\gamma\gamma}}{4}a F^{\mu\nu}\tilde{F}_{\mu\nu}$, but it can be technically natural to set $g_{a\gamma\gamma}=0$ assuming a $Z2$ symmetry of which $a$ and $\gamma$ are different representations. This also sets to zero the kinetic mixing $\mathcal{L}\supset g_{\rm kin}F^{\mu\nu}{F}^{\chi}_{\mu\nu}$, which would also contribute to stellar cooling.} as well as astrophysical bounds due to horizontal branch stars and SN1987a \cite{Raffelt:1996wa}; however, ALPs could still contribute to stellar cooling via plasmon decay $\gamma\rightarrow a + \chi$, which is possible in a medium as the photon dispersion relation allows for such a decay to happen. We will explore these bounds in Section~\ref{starcool}. Notice that these bounds can be avoided if $a$ and $\chi$ are heavy and almost degenerate in mass.
Finally, there is another parameter which can be varied to fit CIBER data, the ALP number density $n_a^{(0)}=R \, \rho_{\rm DM}/m_a$, where $R/m_a$ is a numerical factor and $\rho_{\rm DM}$ is the total DM energy density. 

\subsection{ALP warm dark matter}
In the following we will also consider the scenario in which the ALP population represents a small warm dark matter (WDM) contribution to the DM energy density. This implies an additional fourth tunable parameter, namely the effective temperature. WDM can be produced both thermally or nonthermally \cite{Mazumdar:2016nzr, Shi:1998km}. In the first case, we suppose that the abundance is given by $f_a(\bold{p}_a)=1/[\exp(|\bold{p}_a|/T_{\rm th}(z'))-1]$. 

The distribution could arise, for example, if the ALP and the hidden photon were in thermal equilibrium with the primordial plasma in the early universe; their population would be the result of relativistic decoupling, similar to what happens to neutrinos. The processes which contribute the most to the equilibrium are pair annihilations in the $s$ channel $e^+ + e^-\leftrightarrow a+\chi$ and to a lesser extent plasmon decay $\gamma\leftrightarrow a+\chi$, which is possible in the early universe just as in stars. While plasmon decay is negligible in the early universe production of ALPs, it is very relevant for star cooling, as previously stressed and as we will show in Section~\ref{starcool}. Other processes like $a + e^-\leftrightarrow \chi+ e^-$ could slightly affect the dark sector effective temperature after decoupling, without changing the number density.

The cross section for pair annihilation is
\begin{equation}
\sigma(e^+ + e^-\rightarrow a+\chi)=\frac{\alpha g_{a\chi\gamma}^2}{96} ;
\end{equation}
the process drives $a$ and $\chi$ out of equilibrium when the thermal width is roughly comparable to the Hubble function, viz. $\langle\Gamma_{e^+ e^-}\rangle\simeq H$. 
By noting that in a radiation dominated universe $H=1.66\sqrt{g^*}T^2/m_{\rm Pl}$, where $g^*$ is the effective number of relativistic degrees of freedom at the decoupling and $m_{\rm Pl}=1.22\times10^{19}\rm \, GeV$ is the Planck scale \cite{Kolb:1990vq}, this happens when
\begin{equation}
3\frac{\zeta(3)}{\pi^2}T^3\frac{\alpha g_{a\chi\gamma}^2}{96} \simeq1.66\sqrt{g^*}\frac{T^2}{m_{\rm Pl}}
\end{equation}
which is
\begin{equation}
T_{\rm dec}\simeq 4.8 \times 10^3 \left(\frac{10^{-9} \rm GeV^{-1}}{g_{a\chi\gamma}}\right)^2 \rm GeV \ .
\end{equation}
Let's suppose that there is no new physics between the electroweak phase transition and the decoupling scale. If ALPs decouple at $T_{\rm dec}$, their number density at late times is governed by $g_{*s}$, the
effective number of thermal degrees characterizing the entropy
at the decoupling epoch.The number density of ALPs at low redshift is therefore
\begin{equation}
n_a^{(0)}=\frac{g_{*s}(T^{(0)})}{g_{*s}(T_{\rm dec})}\frac{n_\gamma^{(0)}}{2}
\end{equation}
where the sum is over all the particle content of the standard model plus $a$ and $\chi$. Assuming $g_{a\chi\gamma}\simeq6\times10^{-9} \rm GeV^{-1}$ (i.e. $\tau\simeq9\times 10^{20} \rm s$), we approximately find
\begin{equation}
n_a^{(0)}\simeq\frac{20}{\rm cm^3} \ .
\end{equation}

The most important consequence of a high decoupling temperature is that big bang nucleosynthesis constraints are relaxed. The energy density of the ALPs (one degree of freedom) and of the hidden photons (two or three degrees of freedom depending on the mass) is suppressed compared to the energy density of additional sterile neutrinos or axions which decouple later. To compare, suppose we had introduced a sterile neutrino, which would have implied at least 2 new degrees of freedom. If the decoupling is after the QCD phase transition its contribution to the effective number of neutrinos $\Delta N_{\rm eff}\simeq 0.57$ would have been at least three times larger than $\Delta N_{\rm eff}\lesssim 0.2$ due to $a$ and $\chi$ \cite{Aghanim:2018eyx}. This conclusion remains approximately valid as far as the decoupling takes place before the QCD phase transition.

To recap, there are cosmological constraints coming from measurements of the temperature and polarization power spectra of the CMB anisotropies, the large-scale matter power spectrum, and the Hubble expansion rate \cite{Hannestad:2003ye,Archidiacono:2013cha}, which usually apply to a lately decoupled axion. However, these bounds can be relaxed depending on the decoupling temperature of the ALP. Interestingly, better cosmological measurements could exclude also this class of ALP models.

Alternatively, a nonthermally produced dark matter can have a momentum distribution with a strongly model dependent functional dependence, typical of freeze-in scenarios, where the distribution is set by the thermal distribution of the parent particle, the masses and the couplings \cite{Shaposhnikov:2006xi,Kusenko:2006rh, Petraki:2007gq, Hall:2009bx,Baer:2014eja,Merle:2017jfn}. An out-of-equilibrium decay of heavy particles can also alter the clustering properties of dark matter \cite{Patwardhan:2015kga}. In the case of very low reheating scenarios, cosmological upper bounds on the mass of hot dark matter can be relaxed \cite{deSalas:2015glj}. One can assume a benchmark distribution $f_a(\bold{p}_a)=R_{\rm nth}/[\exp(|\bold{p}_a|/T_{\rm nth}(z'))-1]$, where $R_{\rm nth}$ is again a numerical factor and $T_{\rm nth}^{(0)}$ can be in principle higher than the CMB temperature.
Equation~(\ref{eq:enintensity}) can be expressed in terms of special functions with these distribution. The Dirac delta function can be used to get rid of the angular part of the $\bold{p}_a$, and this would introduce a minimum absolute value of the momentum $|\bold{p}_a^{(0) \rm, min}|$:

\begin{align}
W(z',\omega')&=\frac{R_{\rm nth}}{(2\,\pi)^3}\frac{(1+z)^2}{1+z'}   e^{-\Gamma t(z')}\frac{\Gamma}{H(z')}\frac{m_a}{2 \, {\omega_{\rm max}}} \nonumber \\
&\times\frac{1}{6} \left[3 |\bold{p}_a^{(0) \rm, min}|^2-6 T^2 \text{Li}_2\left(e^{|\bold{p}_a^{(0) \rm, min}|/T}\right)\right.
\nonumber \\
&-6 i \pi  |\bold{p}_a^{(0) \rm, min}|T\nonumber\\&
\left.-6 |\bold{p}_a^{(0) \rm, min}| T \log
   \left(e^{|\bold{p}_a^{(0) \rm, min}|/T}-1\right)+2 \pi ^2 T^2\right]  \ ;
\end{align}
where $\text{Li}_2$ is the polylogarithm of order 2, $T=T^{(0)}_{\rm th}$ (or $T=T^{(0)}_{\rm nth}$) and
\begin{align}
|\bold{p}_a^{(0) \rm, min}|=\pm\frac{ \left[(m_a^2-m_{\chi }^2)^2-4 m_a^2 {\omega'}^2\right]}{4(1+z') {\omega'}^2
   \left(m_a^2-m_{\chi }^2\right)}
\end{align}
depending on whether $\omega'$ is smaller or bigger than $\omega_{\rm max}$.
\\

The photon intensity spectrum due to the astrophysical diffuse, assumed to be \cite{Stecker:2016fsg}, plus the ALP decay contribution is shown in Figures  \ref{fig:alpdecay} and   \ref{fig:alpdecaywarm} for different choices of the parameters. We plot two CDM scenarios with small and large $\omega_{\rm max}$ (model A and B) in Figure \ref{fig:alpdecay}. In Figure \ref{fig:alpdecaywarm} we show two WDM scenario, assuming for one a thermally produced ALP (model $\rm C_{th}$), and for the other an extremely large $T_{\rm nth}$, to make more evident the ALP kinetic energy effect on the photon spectrum (model $\rm C_{nth}$). Considering a thermally produced ALP population (in the following named model $\rm C_{th}$), the intensity spectrum is indistinguishable from the model A CDM spectrum. Model A and $\rm C_{th}$ however differ strongly for what concerns the intensity anisotropies, as we will see below.

\section{Anisotropy constraints}\label{sec:aniso}
\begin{figure*}
        \includegraphics[width=0.32\textwidth]{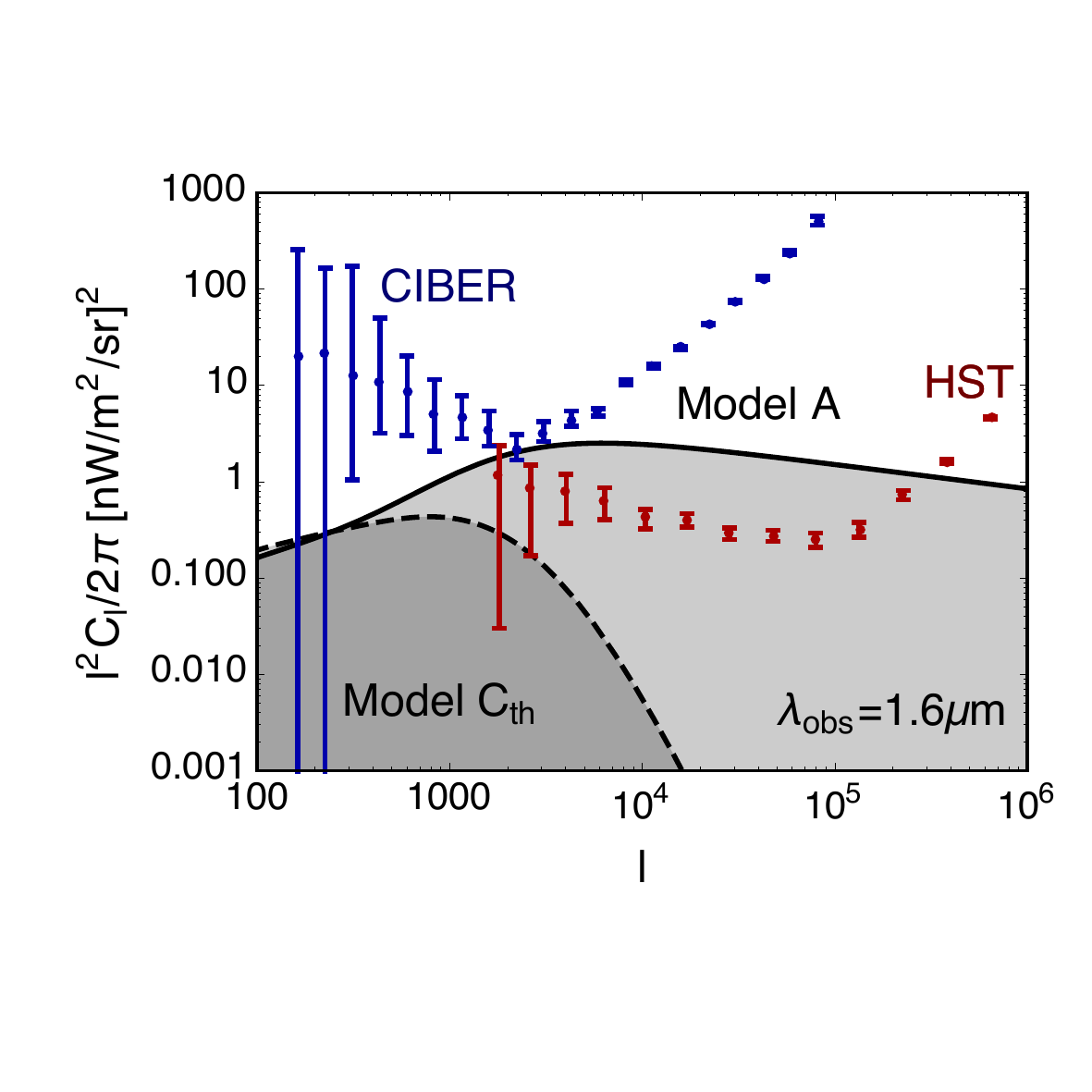}
        \includegraphics[width=0.32\textwidth]{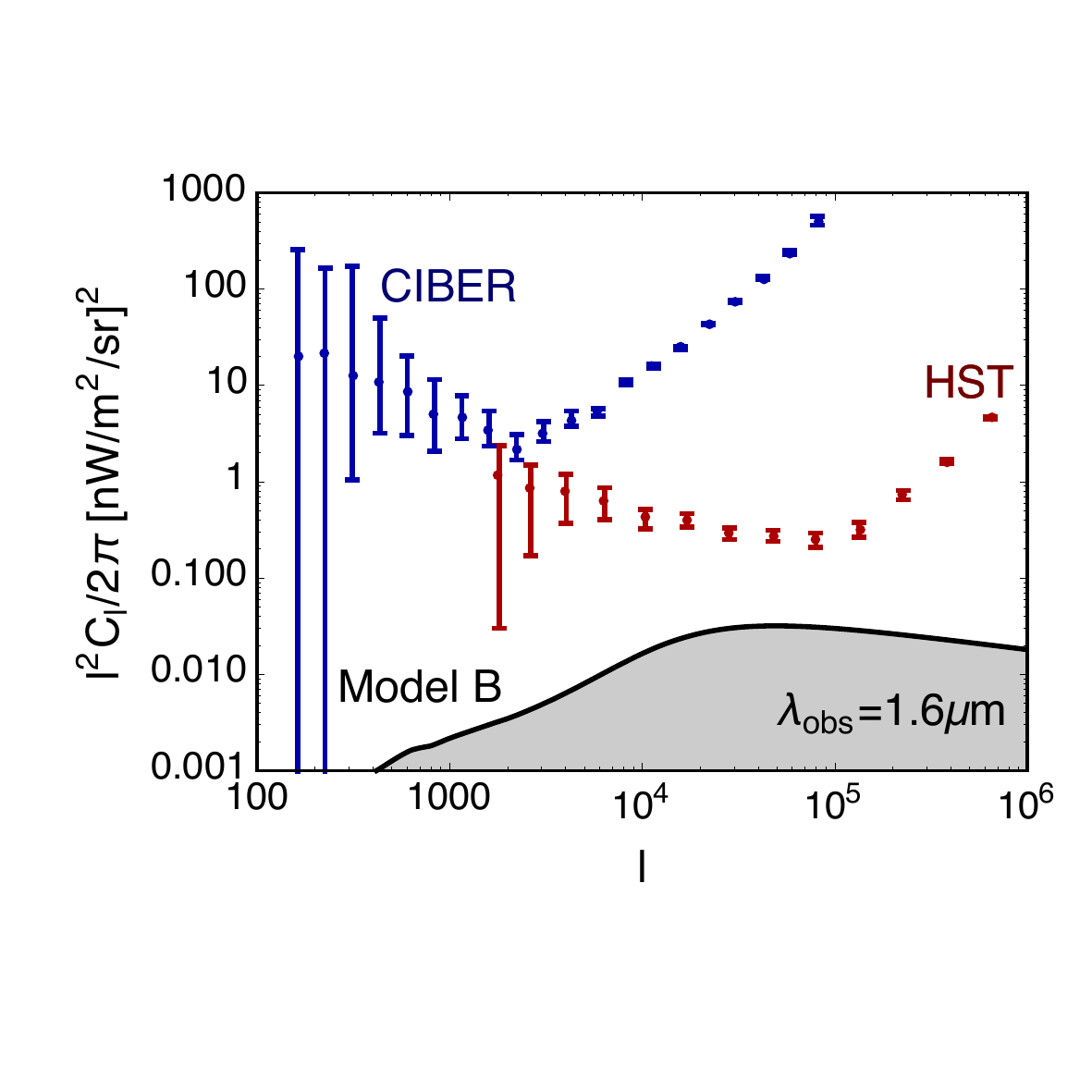}
        \includegraphics[width=0.32\textwidth]{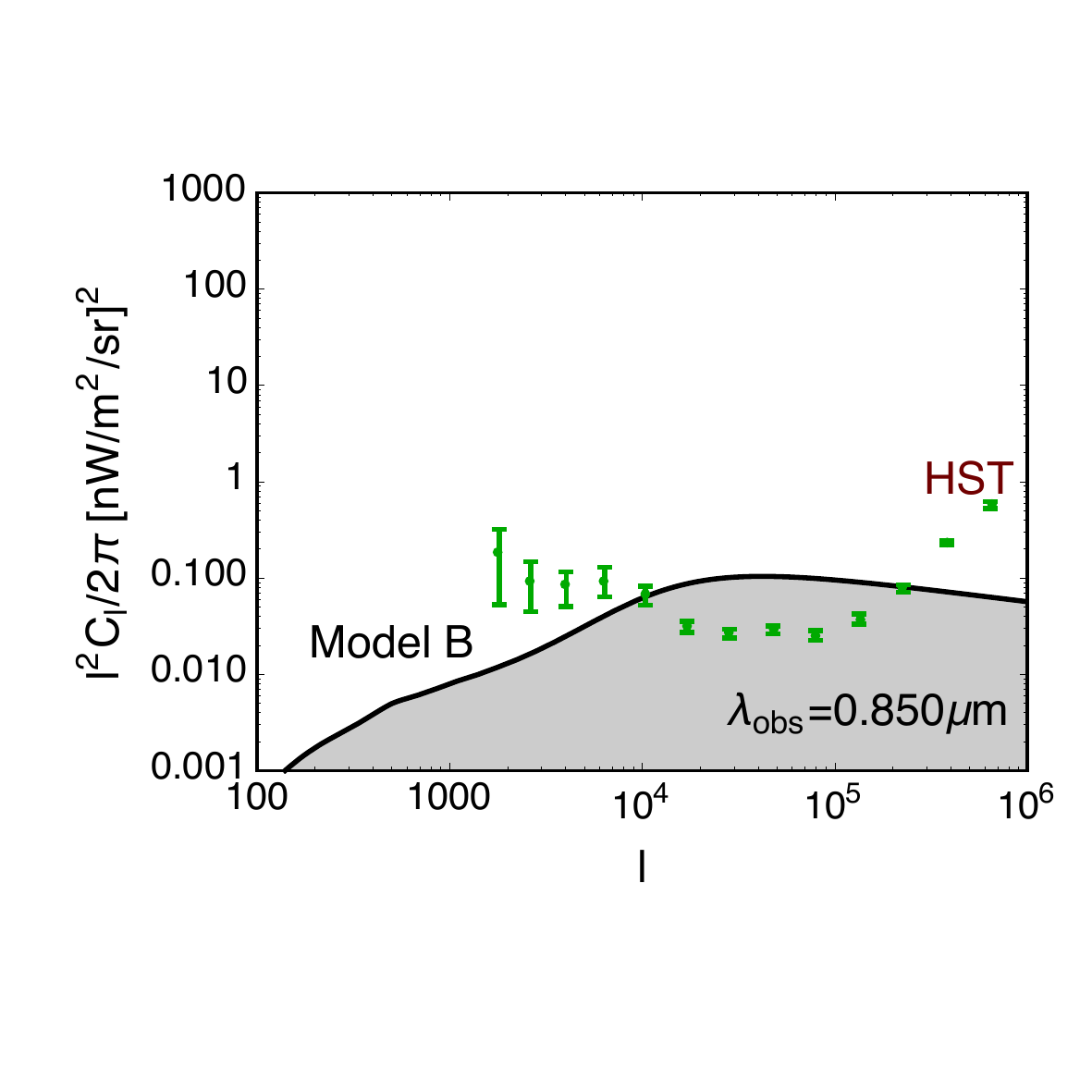}
  \vskip -3em
  \caption{Angular power spectrum due to the decay of an ALP. The data shown are the anisotropies detected by CIBER at observational wavelength $\lambda_{\rm obs}=1.6\, \mu\rm m$ (dark blue), and by HST at observational wavelength $\lambda_{\rm obs}=1.6\, \mu\rm m$ (dark red) and $\lambda_{\rm obs}=0.85\, \mu\rm m$ (dark green). \textit{Left}: anisotropies in the $1.6 \, \mu\rm m$ band for models A (solid line) and $\rm C_{th}$ (dashed line); \textit{center}: anisotropies in the $1.6 \, \mu\rm m$ band for model B; \textit{right}: anisotropies in the $0.85 \, \mu\rm m$ band for model B.}
  \label{fig:alpanisotropy}
\end{figure*}

The gravitational clustering of dark matter makes the photon flux produced by the decaying ALP anisotropic. In this section we revisit the calculations as done in \cite{Kohri:2017oqn,Gong:2015hke}. We take into account the energy resolution of the detector,  following \cite{Fernandez:2009fk}. To the best of our knowledge, this is the first time such a procedure is carried out in the context of infrared photons produced by dark matter decay. Our main goal in this section is then to revisit the bounds on CDM scenarios from anisotropies, and to extend the analysis to a WDM scenario.

The average intensity of the flux detected in an energy band centered in $\omega$ with width $\Delta\omega$ is given by
\begin{align}
I(\omega,\Delta\omega)=\frac{1}{\Delta\omega}\int_{\Delta\omega} d\omega \, \omega^2 \int_z^{\infty}dz' W(z',\omega')
\end{align}
assuming a $\Delta \omega=\omega$ flat passband filter for the detector \cite{Bock:2012fw}; the fluctuations toward a direction of the sky $\hat{\bold{n}}$ can be expanded as spherical harmonics
\begin{align}
\delta I(\omega,\Delta\omega,\hat{\bold{n}})=&I(\omega,\Delta\omega,\hat{\bold{n}})-I(\omega,\Delta\omega)\nonumber\\
&=\sum_{l,m}a_{l,m}(\omega,\Delta\omega)Y_{l,m}(\hat{\bold{n}}) \ .
\end{align}

Anisotropies are often conveniently described in terms of the angular power spectrum (suppressing the $\omega$ dependence from the notation)
\begin{equation}
C_l(\Delta\omega)=\langle|a_{l,m}(\Delta\omega)|^2\rangle=\frac{1}{2l+1}\sum_{m=-l,+l}|a_{l,m}(\Delta\omega)|^2
\end{equation}
which written in terms of the window function is
\begin{align}
C_l(\Delta\omega)&=\frac{1}{\Delta\omega}\int_{\Delta\omega} d\omega_1 \, \omega_1^2 \int_z^{\infty}dz_1' W(z'_1,\omega')\nonumber\\
&\times\frac{1}{\Delta\omega}\int_{\Delta\omega} d\omega_2 \, \omega^2_2 \int_z^{\infty}dz_2' W(z'_2,\omega'_2)\nonumber\\&
\times\frac{2}{\pi}\int dk k^2 P_\delta\left(k,r(z'_1),r(z'_2))\right.
\nonumber\\&
\qquad \qquad \qquad \qquad \left.j_l(kr(z'_1))j_l(kr(z'_2)\right)
\end{align}
where $r(z)=\int_0^z dz/H(z)$ is the comoving distance, $j_l(kr(z))$ is the spherical Bessel function and the power spectrum (i.e., the density contrast) is defined as $\langle\delta_{\bold{k}_1}(r(z_1))\delta_{\bold{k}_2}(r(z_2))\rangle=(2\pi)^3\delta^{(3)}(k_1-k_2)P_\delta(k_1,r(z_1),r(z_2))$. If the power spectrum varies slowly as a function of $k$ we can use Limber approximation \cite{Ando:2005xg}, which is correct up to $\mathcal{O}(l^{-2})$ \cite{LoVerde:2008re}
\begin{align}
\frac{2}{\pi}\int dk k^2 P_\delta(k,r(z'_1),r(z'_2))j_l(kr(z'_1))j_l(kr(z'_2))\nonumber\\
\simeq\frac{1}{r(z_1')^2}  P_\delta\left(k=\frac{l}{r(z_1')},r(z'_1)\right)\delta^{(1)}(r(z_1')-r(z_2')) .
\end{align}
Notice that we do not have to worry about the sharpness of the differential flux caused by the delta function in the window function for CDM, because this is cured by averaging over the energy bandwidth of the detector. This procedure has been used for similar analyses with gamma rays (see e.g. \cite{Ibarra:2009nw,Ando:2013ff}). Defining $z_{\rm M}=\omega^{\rm max}/(\omega-\Delta\omega/2)-1$ and $z_{\rm m}=\omega^{\rm max}/(\omega+\Delta\omega/2)-1$ as the maximum and minimum redshift observed in the anisotropy measurement, we have
\begin{align}
C_l(\Delta\omega)=&\int_{z_{\rm m}}^{z_{\rm M}}dz\left\{\frac{1}{4\pi}\frac{e^{-\Gamma t(z)}}{H(z)(1+z)^3}\omega_{\rm max}^2\Gamma n^{(0)}_a \frac{1}{\Delta\omega}\right\}^2\nonumber\\
&\times \frac{1}{r(z)^2}  P_\delta\left(k=\frac{l}{r(z)},r(z)\right)H(z) \ .
\end{align}
Our redshift dependence agrees with the one of Equation (A10) of \cite{Fernandez:2009fk}, because we are considering the angular power spectrum of the energy flux (units are energy \textit{squared} per time, per surface, per steradians and per energy); to compare the results of \cite{Fernandez:2009fk} and ours, Equation (A1) of the same reference shall be multiplied times $\nu$, which gives an additional $(1+z)^{-2}$ in the final expression.

The anisotropy power spectra for lighter ($\omega_{\rm max}=1\, \rm eV$) and heavier ($\omega_{\rm max}=8\, \rm eV$) dark matter are shown in Figure \ref{fig:alpanisotropy}, where they are compared with data of CIBER \cite{Zemcov:2014eca} and of the Hubble Space Telescope (HST) \cite{Mitchell-Wynne:2015rha}. The matter power spectrum has been calculated through CLASS code \cite{Blas:2011rf}, publicly available at \cite{Lesgoursite}. In the first case, we explored both the CDM and the WDM cases (assuming $m_a=2 \, \rm{eV}$ for the latter case). The WDM power spectrum has been computed in the adiabatic approximation \cite{Inman:2016qmg}, $P_{\delta, \rm WDM}=(\mathcal{T}_{\rm WDM}/\mathcal{T}_{\rm CDM})^2P_{\delta, \rm CDM}$, where $\mathcal{T}$ is the transfer function.\footnote{The transfer function $\mathcal{T}_{\rm WDM}$ must be evaluated including the dominant dark matter component, which is assumed here cold and different from the decaying ALP.} The latter relates the primordial and the present day power spectra \cite{Eisenstein:1997jh}, and is another CLASS output \cite{Lesgourgues:2011rh}. In all cases, given that we needed to integrate over the redshift, we assumed conservatively a linear evolution for the matter power spectrum, using the non-linear matter power spectrum $P_\delta$ obtained with CLASS, calculated at redshift $z=0$ and evolved backwards
\begin{align}
P_{\delta}\left(k=\frac{l}{r(z)},r(z)\right)=P_{\delta}\left(k=\frac{l}{r(z)},r=0\right)D(z)^2 \ .
\end{align}
Here, $D(z)\propto H(z)\int_z^\infty dz' (1+z')H(z')^{-3}$ is the linear growth factor, to be normalized with $D(0)=1$ \cite{Ibarra:2009nw}.

As heuristically expected, WDM evades quite easily the constraints due to anisotropy measurements, as understood by showing the model $\rm C_{th}$ anisotropy spectrum (dashed line in left panel of Figure~\ref{fig:alpanisotropy}). These become unrestrictive when considering a non-thermally produced hot dark matter with high effective temperature, like in model $\rm C_{nth}$, as their free-streaming length is even larger. Light CDM (model A) can be considered excluded by our analysis.

For what concerns heavier dark matter (model B), our results are shown in the central and right panels of Figure \ref{fig:alpanisotropy}, where the anisotropy power spectrum is computed both for the $1.6\,\mu \rm{m}$ wavelength band (light red) and for the $0.85\,\mu \rm{m}$ band. The $0.85\,\mu \rm{m}$ band slightly overshoots the observed data in the relevant wavelength; however, the exclusion is much weaker than what has been found in previous analysis \cite{Kohri:2017oqn}, due to averaging over the detector bandwidth.

A final comment is required about the anisotropy measurements. While our goal in this section has been to revisit previous analyses accounting for the detector bandwidth, a cold dark matter origin for the CIBER excess is still excluded, even if less strongly than previously thought. On the other hand, a thermal ALP population origin is not falsified by anisotropy measurements. Nevertheless, anisotropies hint either to the possible presence of an additional astrophysical class of sources to the EBL, which would possibly explain the angular power spectra of difference wavelenghts complementing the dominant contributions of shot power at low multipoles and galaxies at high multipoles \cite{Gong:2015hke}, or to a different modelling of the latter.
\section{Star cooling constraints}
\label{starcool}
The processes by which the populations of ALPs $a$ and hidden photons $\chi$ are mostly produced in a plasma depend on the temperature and density conditions of the stars considered. Let us consider plasmon decay $\gamma\rightarrow~a+\chi$. Other processes like photo-production, pair annihilation of photons or bremsstrahlung are suppressed by a higher order in the coupling $g_{a\chi\gamma}$ or $e=\sqrt{4\pi\alpha}$. While these processes can be relevant for other kind of particles and interactions, we anticipate that the strongest constraints come from stars which would mostly emit $a$ and $\chi$ through plasmon decay. 

Given that we are interested in an order of magnitude estimate, we will not take into account the longitudinal plasmon decay, as it would be a negligible correction, keeping only the transverse plasmon decay into account \cite{Raffelt:1996wa}. The longitudinal plasmon decay in fact contributes negligibly to the cooling, because there is no resonant conversion from longitudinal plasmon to pseudoscalars \cite{Hardy:2016kme}. The decay of a strongly nonrelativistic plasmon is due to the coupling
\begin{equation}
\mathcal{L}\supset\frac{g_{a\chi\gamma}}{4}a F^{\mu\nu}\tilde{F}^\chi_{\mu\nu}=\frac{g_{a\chi\gamma}}{2}a\bold{E}\cdot\bold{B}^\chi \ ,
\end{equation}
because the oscillation of the plasma is purely electric when the momentum of the plasmon is much smaller than its frequency. In general, $\bold{E}= -\nabla A^0 -\partial_t \bold{A}$ and $\bold{B}=\nabla \times \bold{A}$, so that $\bold{E}\propto \omega \bold{\epsilon}_T$ and $\bold{B}^\chi\propto \bold{k}_\chi \times \bold{\epsilon}_\chi$, where $\omega=\sqrt{\bold{k}^2+m_T^2}\simeq m_T$ is the frequency of the plasmon. The usual Feynman diagram rules then give
\begin{align}
\overline{|\mathcal{M}|^2}&=\frac{g_{a\chi\gamma}^2}{4}\frac{m_T^2}{2}\sum_{\epsilon_T, \epsilon_\chi}\left| \bold{\epsilon}_T\cdot(\bold{k}_\chi \times \bold{\epsilon}_\chi)\right|^2\\&=\frac{g_{a\chi\gamma}^2}{4}\frac{m_T^2}{2}\sum_{\epsilon_T}\left| \bold{\epsilon}_T\times\bold{k}_\chi \right|^2 
\end{align}

where $m_T$ is the ``transverse photon mass''. The sum over the transverse polarizations is in the Coulomb gauge
\begin{equation}
\sum \epsilon_T^i \epsilon_T^j=\delta^{i}\delta^j-\frac{k^i k^j}{|\bold{k}|^2}
\end{equation}
which gives
\begin{equation}
\overline{|\mathcal{M}|^2}=\frac{g_{a\chi\gamma}^2}{32}m_T^4(1+\cos^2\theta)
\end{equation}
where $\theta$ is the angle between the plasmon and the hidden photon momenta. Including a boost factor $\frac{m_T}{\omega}$, the decay rate of a plasmon with frequency $\omega$ is given by 
\begin{equation}\label{plsrate}
\Gamma_\gamma=\frac{1}{3}Z_T\frac{g_{a\chi\gamma}^2}{128\pi}m_T^3\frac{m_T}{\omega}
\end{equation}
where $Z_T$ is the vertex renormalization funcion \cite{Raffelt:1996wa}. This expression reduces to
\begin{equation}
\Gamma_\gamma=\frac{1}{3}\frac{g_{a\chi\gamma}^2}{128\pi}\omega_{\rm{pl}}^3\frac{\omega_{\rm{pl}}}{\omega}
\end{equation}
in the nonrelativistic, nondegenerate limit of the plasma. This is the formula to be used in most of stellar plasma cases, where $m_T\simeq\omega_{\rm pl}$ and
\begin{equation}
\omega_{\rm pl}^2=\frac{4\pi\alpha n_e}{m_e},
\end{equation}
with $n_e$ electron number density and $m_e$ electron mass \cite{Raffelt:1996wa}. Notice that our result is $1/3$ smaller than the one found in \cite{Kohri:2017oqn}.\footnote{Using equation \eqref{plsrate} for relativistic plasma,  $m_T\simeq\thinspace 3 \,\omega_{\rm pl}/2$ and $\omega_{\rm pl}^2=\frac{4\pi\alpha T^2}{9}$, one sees that in the early universe plasmon decay is negligible compared to pair annihilation.}

The production of ALPs and hidden photons in horizontal branch stars through plasmon decay (when their mass is smaller than the plasma frequency) puts bounds on the coupling $g_{a\chi\gamma}$. The energy loss per unit mass due to plasmon decay is given by 
\begin{equation}
\epsilon=\frac{1}{\rho_s \pi^2}\int dk k^2 \frac{\omega}{e^{\omega/T}-1}\Gamma_\gamma
\end{equation}
where $\rho_s$ is the mass density of the star. So we obtain
\begin{align}
&\epsilon=\frac{\zeta(3)}{192\pi^3}\frac{\omega_{\rm pl}^4 T^3 g_{a\chi\gamma}^2}{\rho_s}\simeq 0.6 \rm{\, erg/g/s}\nonumber \\&\times \left(\frac{\omega_{\rm {pl}}}{1 \rm{\,keV}}\right)^4\left(\frac{T}{10 \rm{\,keV}}\right)^3 \left(\frac{10^4 \rm{\,g/cm^3}}{\rho_s}\right) \left(\frac{g_{a\chi\gamma}}{10^{-8} \rm{\, GeV}^{-1}}\right)^2
\end{align}
with canonical parameters of horizontal branch star cores; the star cooling bound implies that $\epsilon~\lesssim~10~\rm{\, erg/g/s}$. 
A more stringent bound is given by the required agreement between the
predicted and observationally inferred core mass at the helium flash of red giants.
This is to be expected, since the bounds on the coupling $g_{a\chi\gamma}$ can be directly read from the existing constraints on a putative neutrino magnetic dipole moment $\mu_\nu$. The plasmon decay rate is the same for both channels \cite{Raffelt:1996wa}, after substituting
\begin{equation}
g_{a\chi\gamma}\rightarrow 4 \mu_\nu \ ;
\end{equation}
non-standard neutrino losses would delay the ignition of helium in low-mass red giants \cite{Raffelt:1999gv}. With a 95\% confidence level $\mu_\nu\lesssim 1.4\times10^{-9}\rm GeV^{-1}$ \cite{Viaux:2013lha}, which translates to the bound $g_{a\chi\gamma}\lesssim 6\times10^{-9}\rm GeV^{-1}$. Interestingly, a cooling excess has been claimed for this class of stars \cite{Giannotti:2015kwo}, and the plasmon decay to an ALP and a hidden photon with a coupling of this size would contribute as an additional cooling channel. It shall be noted however that plasmon decay cannot account for some of the cooling hints \cite{Giannotti:2015kwo,Giannotti:2017hny}.

\section{Gamma-ray attenuation 
}\label{sec:atten}

The increased EBL flux has observable impact on the propagation of very high energy $E>0.1$~TeV photons due to enhanced rate of $e^{+}e^{-}$ pair production process. This effect may relax the tension between the predicted $\gamma$-ray flux and the Fermi LAT measurement of isotropic gamma-ray background (IGRB)~\cite{Ackermann:2014usa} in traditional multimessenger scenarios of high energy neutrino origin (see e.g.~\cite{Murase:2013rfa, Tamborra:2014xia, Ando:2015bva, Bechtol:2015uqb, Giacinti:2015pya,Gao:2016uld}) and eliminate need of hidden cosmic-ray accelerator~\cite{Murase:2015xka}.
\begin{figure*}
	\centering
		\includegraphics[width=0.55\textwidth]{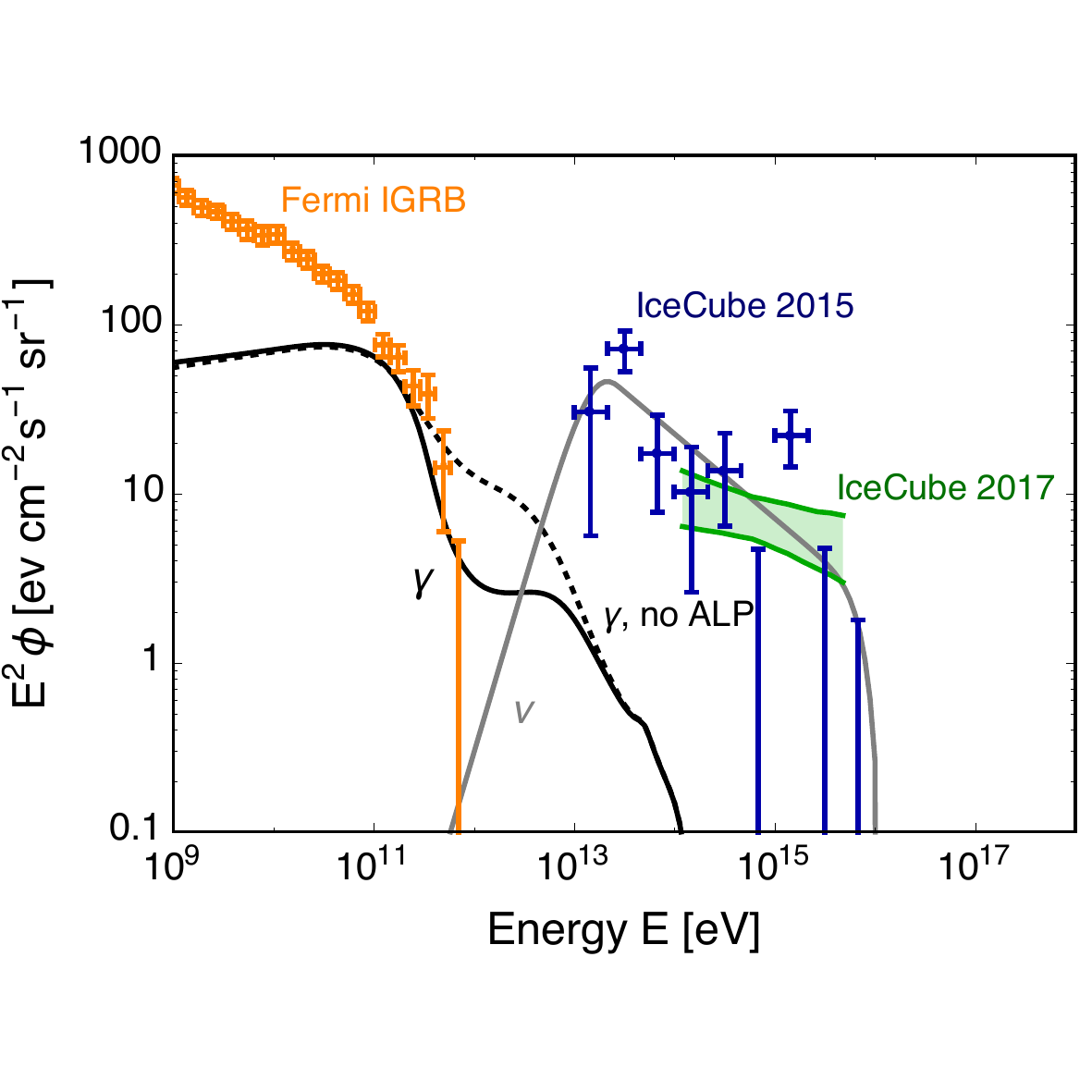}
		  \vskip -3em
	\caption{The $\gamma$-ray and neutrino fluxes expected in a minimal $p\gamma$ production scenario of Ref.~\cite{Murase:2015xka} (see details in text). 
Also shown are the per-flavor IceCube neutrino flux according to~\cite{Aartsen:2015knd} (blue error bars) and more resent estimate~\cite{Aartsen:2017mau} (green band).   The gamma-ray flux in the absence of ALP decays (dotted line) is decreased in the presence of an additional EBL component (solid line), which alleviates the tension with 
Fermi LAT IGRB measurements~\cite{Ackermann:2014usa}.
}
	\label{fig:pgamma}
\end{figure*}
In Fig.~\ref{fig:pgamma} we illustrate the effect. We calculate the  neutrino and the accompanying $\gamma$-ray flux in the minimal $p\gamma$ production scenario of Ref.~\cite{Murase:2015xka} with $\epsilon_{\nu}^b=25$ TeV, assuming low X-ray luminosity AGN evolution of Ref.~\cite{Hasinger:2005sb} for the sources and the minimal EBL model~\cite{Stecker:2016fsg} with or without the contribution from ALP, for which we use model A.\footnote{We remark again that model A and $\rm C_{th}$ are indistinguishable at the level of the intensity spectrum.} The spectra shown were obtained by solving transport equations for neutrinos and electron-photon cascades with the public numerical code~\cite{Kalashev:2014xna}. The effect of the increased EBL is clearly seen on the $\gamma$-ray flux above 100 TeV.
In principle, the enhanced Universe opaqueness for $\gamma$-rays predicted in the above scenario will only sharpen the well known problem of unexpectedly hard $\gamma$-ray spectra detected from the remote blazars. In Appendix~\ref{AppB} a consistency check is carried out to verify the compatibility of our scenario with blazar observations. We found that the only parameters range excluded by analysis of the deabsorbed spectra is the one of model B, which is already excluded by the observed angular power spectrum, whereas models A and $\rm C_{th (nth)}$ are viable.

\section{Conclusions}
\label{conc}
 
In this paper we have explored the possibility that the high EBL spectrum detected by the CIBER collaboration could be due to the decay of an axionlike particle with mass around an electronvolt. Taking into account multimessenger, multiwavelength observations, we have shown that a warm dark matter component, produced either thermally or nonthermally, can explain the enhanced EBL detected by the sounding rocket CIBER.  The increased level of EBL alleviates the tension between the neutrino flux detected at IceCube and the gamma-ray flux measured by Fermi, assuming a p$\gamma$ production scenario. We have shown that the anisotropy measurements do not exclude this solution, and we have studied the effect on the propagation of $\gamma$ rays detected from distant sources, such as the Blazar Lac PG 1553+113. The ALP we consider is not in contradiction with current astrophysical observations, and  the concordance of multimessenger, multiwavelength data lends credibility to the hypothesis that a decaying particle contributes to the measured excess of infrared background radiation.

\begin{acknowledgments}
We thank Y.~Gong and M.~Zemcov, as well as K.~Nakayama, for helpful discussions about the anisotropy power spectrum. We also thank J.~Redondo, A.~Millar and G.~Raffelt for useful comments and discussions. E.V. thanks J.~Stadler for introducing him to CLASS. E.V. thanks for the hospitality the Kavli IPMU, where part of this work has been carried out, and acknowledges support by the European Union through Grant No. H2020-MSCA-ITN-2015/674896 (Innovative Training Network ``Elusives''). O.K. acknowledges support by the Foundation for the Advancement of 
Theoretical Physics and Mathematics ``BASIS'' grant 17-12-205-1.  A.K. was supported by the U.S. Department of Energy Grant No. DE-SC0009937, and by the World Premier International Research Center Initiative (WPI), MEXT, Japan.   Part of this work was performed at Aspen Center for Physics, which is supported by National Science Foundation grant PHY-1607611.
\end{acknowledgments}

\begin{figure*}\centering
		\includegraphics[width=0.46\textwidth]{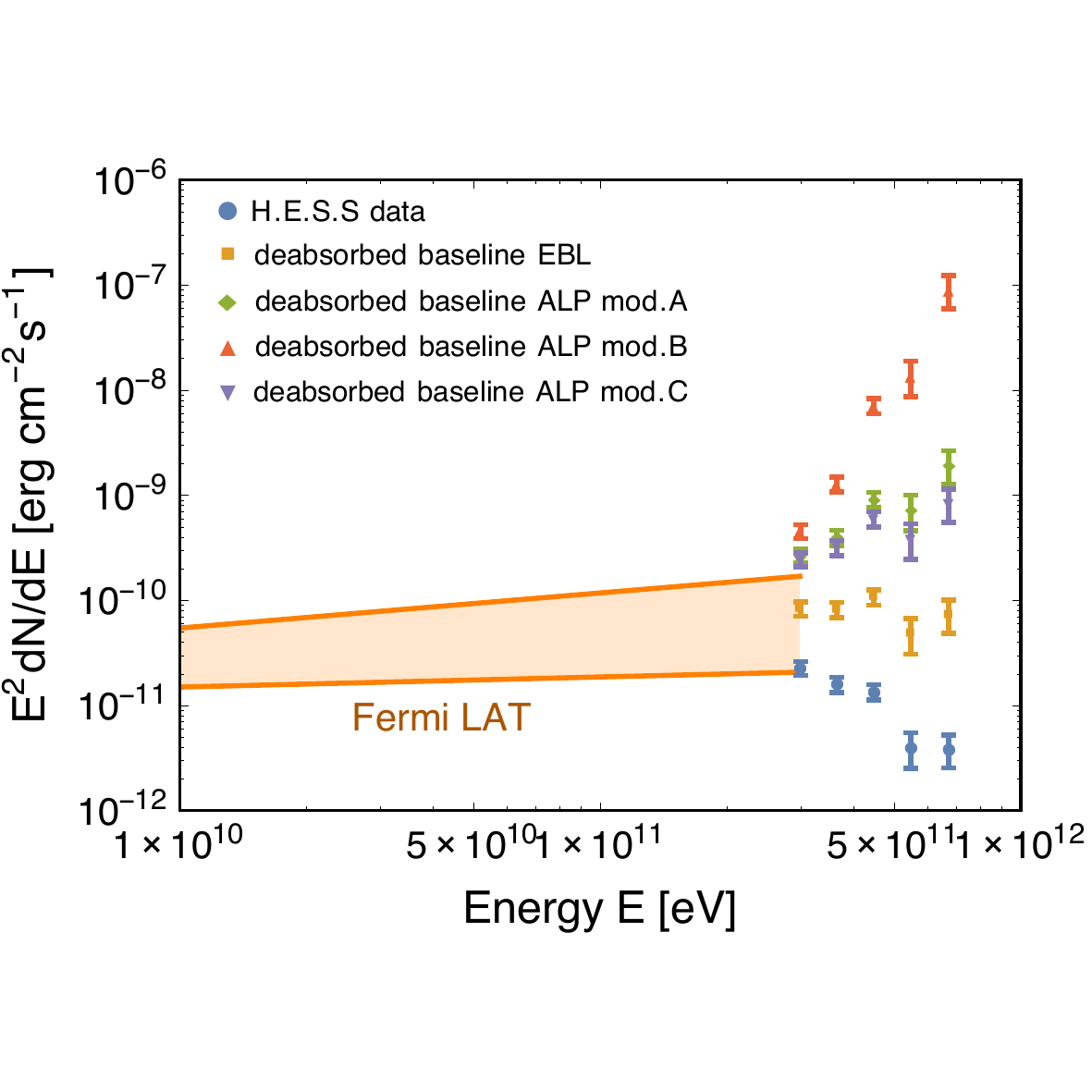}
				  \vskip -3em
	\caption{Observed spectral energy distribution of PG 1553+113 measured during the flare by Fermi LAT (power law approximation) and H.E.S.S. (as shown in Fig.3 of Ref.~\cite{Abramowski:2015ixa}) together with the deabsorbed spectra calculated using EBL model  of Ref.~\cite{Inoue:2012bk} with or without extra contribution from ALP. 
}
	\label{fig:de-absorbed}
\end{figure*}
\appendix

\section{Consistency with blazar observables}
\label{AppB}
The enhanced Universe opaqueness for $\gamma$-rays predicted in the ALP decay scenario sharpens the well-known problem of unexpectedly hard $\gamma$-ray spectra detected from the remote blazars. A possible solution to this problem proposed in Ref.~\cite{Essey:2009zg,Essey:2009ju,Essey:2010er,Essey:2011wv,Murase:2011cy,Prosekin:2012ne,Aharonian:2012fu,Kalashev:2013vba,Inoue:2013vpa,DeFranco:2017wdr} is based on the natural assumption that the blazars also emit ultrahigh energy cosmic rays which contribute to the observed $\gamma$-ray flux through secondary electromagnetic cascades produced in line of sight cosmic ray interactions. The above scenario allows to avoid exponential $\gamma$-ray flux suppression with distance from the source.

The straightforward way to find if an extra component is needed to fit the observations is to construct the so called deabsorbed spectrum, i.e. the primary spectrum recovered from the observations assuming no extra components. The negative break in the deabsorbed spectrum can be considered as a good indication of extra component presence. By definition the deabsorbed spectrum
\begin{equation}
F_{\rm de-absorbed}=\exp(\tau(z,E))F_{\rm observed}
\label{eq:deab}
\end{equation}
depends not only on source redshift but also on the  EBL model assumed through optical depth $\tau$. We will illustrate this point on high-frequency peaked BL Lac object PG 1553+113, one of the most variable remote sub-TeV $\gamma$-ray sources known today. Its $\gamma$-ray flaring activity has been detected by H.E.S.S. telescopes  during the nights of 2012 April 26 and 27 when the source flux above 0.3 TeV increased by a factor of 3 with evident signs of variability on scale of hours~\cite{Abramowski:2015ixa}. In Fig.~\ref{fig:de-absorbed} (left) we show the average spectrum of the object measured during the flare by Fermi LAT and H.E.S.S. (as shown in Fig. 3 of Ref.~\cite{Abramowski:2015ixa}) together with the deabsorbed spectra calculated using EBL model of Ref.~\cite{Inoue:2012bk} with or without extra contribution from ALP decay models A, B and $\rm C_{th(nth)}$. We use lower limit $z>0.43$~\cite{Danforth:2010vy} as a conservative source redshift estimate. It is now clear from the figure that increased EBL may lead to negative break in the deabsorbed spectrum, which indicates presence of extra component.

Let us assume now that the extra component is not as highly variable as we would expect in the case of secondary $\gamma$ from cosmic rays. Would it contradict observations? To answer this question in a conservative manner, we calculate the maximal expected integral flux of primary $\gamma$ above 0.3 TeV during the flare phase $F^{\rm var}_{\rm max}$ and the minimal required integral flux of the constant extra component $F^{\rm ext}_{\rm min}$. We calculate $F^{\rm var}_{\rm max}$ assuming power low injection and maximal initial $\gamma$ flux consistent with Fermi LAT observations below 30 GeV. $F^{\rm ext}_{\rm min}$ is then calculated simply by subtraction of the primary component from the average observed flux at flare phase. For ALP models A, B and C we get $F^{\rm var}_{\rm max}/F^{\rm ext}_{\rm min}$ integral flux ratio equal to 2.3, 0.36 and 7.6 respectively. From the observation that average integral flare flux above 0.3 TeV is 3 times higher than pre-flare flux we infer
\begin{equation}
3=\frac{F^{\rm var}+F^{\rm ext}}{F^{\rm const}+F^{\rm ext}}<\frac{F^{\rm var}+F^{\rm ext}}{F^{\rm ext}}=\frac{F^{\rm var}}{F^{\rm ext}}+1,
\label{eq:flare_flux}
\end{equation}
where $F^{\rm const}$  is possible contribution of primary photons in pre-flare flux. Now it is obvious that the condition $F^{\rm var}/F^{\rm ext}<0.36$ which we have in case of model B contradicts~(\ref{eq:flare_flux}), while other models are still in line with the inequality.

\bibliographystyle{apsrev4-1}
\bibliography{cosmic_paper.bib}

\end{document}